\documentclass[11pt]{article}
\usepackage{vietnam,epsfig}

\def\be{\begin{equation}}
\def\ee{\end{equation}}
\def\bea{\begin{eqnarray}}
\def\eea{\end{eqnarray}}

\begin{document}
\title{Double proton tagging at the LHC as a means to discover new physics}

\author{ B. E. Cox }

\address{School of Physics and Astronomy, The University of Manchester, Oxford Road, Manchester, M139PL, UK}

\maketitle\abstracts{
We review the theoretical and experimental motivations behind recent proposals to add forward proton 
tagging detectors to the LHC experiments as a means to search for new physics.  }

\section{Introduction}

There has been increasing interest in the past few 
years in the possibility of using diffractive interactions as a search tool for new physics. 
In particular, it has been suggested that the so-called central exclusive production process might provide a 
particularly clean environment to search for, and identify the nature of, new particles at the LHC.
By central exclusive, we refer to the process $PP\rightarrow P \oplus \phi \oplus P$, where 
$\oplus$ denotes the absence of hadronic activity ('gap') between the outgoing protons and the 
decay products of the central system $\phi$. An example would be standard model Higgs Boson production, 
where the central system would consist of 2 b-quark jets, and {\it no other activity}. 

The process is attractive for two main reasons. Firstly, if the outgoing protons remain intact and scatter
 through small angles, then, to a very good approximation, the central system 
$\phi$ must be produced in a spin $0$,
 CP even state, therefore allowing a clean determination of the quantum numbers
 of any observed resonance. Secondly, as a result of these quantum number selection 
rules, coupled with the (in principle) excellent mass resolution on the central
 system achievable if suitable proton detectors are installed, signal to 
background ratios of greater than unity are predicted for Standard Model 
Higgs production \cite{DeRoeck:2002hk}, and significantly larger for the 
lightest Higgs boson in certain regions of the MSSM parameter space 
\cite{Kaidalov:2003ys}. Simply stated, the reason for these large signal to 
background ratios is that exclusive $b$ quark production, the primary 
background in light Higgs searches, is heavily suppressed due to the quantum 
number selection rules. Another attractive feature is the ability to directly 
probe the CP structure of the  Higgs sector by measuring azimuthal asymmetries 
in the tagged protons (a measurement previously proposed only at a future 
linear collider) \cite{Khoze:2004rc}.

Given the apparent benefits of the central exclusive process, there 
is an increasing amount of R\&D work aimed at assessing whether it is 
possible to install forward proton detectors with appropriate acceptance 
at ATLAS and / or CMS, and fully integrate such detectors within the 
experimental trigger frameworks. 

\section{Predictions for central exclusive production at the LHC}

 It is the claim of Khoze et. al. that the central exclusive process is 
perturbatively calculable, up to 
the un-integrated off-diagonal parton distributions of the proton 
(oduPDFs) \cite{Khoze:2001xm}, 
and the so-called gap survival factor
which accounts for the probability that there are no interactions between 
the spectator partons in the protons, which would 
destroy the protons and the gaps.    
Sudakov factors, which enter via the requirement that there be no radiation 
into the final state, render the cross section 
calculation perturbative (at least for Standard Model Higgs boson masses).  
This is to be contrasted with earlier approaches in which an infra-red 
cut-off was introduced 'by hand' and tuned to fit the 
total PP cross section \cite{Cudell:1995ki,Levin:1999qu}. The earliest 
calculation of the process was carried out by 
Bialas and Landshoff \cite{Bialas:1991wj}, and has recently been updated 
by Boonekamp et. al.
\cite{Boonekamp:2003wm,Boonekamp:2004nu}. The early predictions for the 
exclusive standard model Higgs production 
cross section at 14 TeV were
 all extremely large (over 100 fb). The predictions of Khoze et. al. are 
orders of magnitude lower.
Fortunately, the calculations (approximately) factorise into a luminosity 
function, which contains the physics of the
colour-singlet gluons, and a hard sub-process cross section. It is therefore 
possible to check the calculations by observing 
the exclusive production of other higher rate processes. The search for 
exclusive $\chi_C$ meson production at the Tevatron, for example, is 
underway at the time of writing, although no results have yet been published.  

According to Khoze et. al. the cross section prediction for the production 
of a 120 GeV standard model Higgs at 14 TeV is 
3 fb \cite{Khoze:2001xm} \footnote{for a fuller review of the uncertainties 
in this calculation, see \cite{Cox:2004rv} and references therein}.
We take this prediction as the benchmark result. For 30fb$^{-1}$ of LHC 
luminosity, therefore, one would expect $\sim 90$
signal events. This is a small number, so the viability of detection depends 
crucially on the acceptance of 
the proton detectors, the efficiency of the trigger, and the magnitude of the 
background. De Roeck et. al.
 have made a detailed study, including calculations of the $b \bar b$ 
backgrounds, the $b$-tagging 
efficiency and the acceptance and mass resolution of possible proton tagging detectors at LHC. 
$b$-tagging is necessary because the exclusive production of gluon jets is not suppressed and therefore has an extremely 
large rate which would totally swamp the Higgs signal.
The bottom line is that, for a luminosity 
of 30 fb$^{-1}$, De Roeck et. al. expect 11 signal events over 3 background. Details can be found in \cite{DeRoeck:2002hk},
but we make a few remarks here. The very low $b \bar b$ backgrounds are a result, as mentioned in the introduction, of the 
spin selection rules which are a consequence of the colour-singlet configuration of the exchanged gluons 
(and strictly the small 
transverse momenta of the outgoing protons). These selection rules are not exact: in fact the $b \bar b$ background is 
proportional to $m_b^2 / E_T^2$, where $E_T$ is the transverse energy of the $b$ jets (which will be of order $m_H/2$).
This is a small effect for a 120 GeV Higgs, but as we shall see, can be important for lighter Higgs bosons which might occur
in certain regions of the MSSM parameter space. The selection rules can also be violated by higher order gluon emission. 
De Roeck et. al. 
consider the contributions from NLO and NNLO diagrams, and find that they are able to reduce them significantly by using a 
combination of the independent mass measurements from the proton taggers and the central detectors. This result can (very crudely) be
pictured as the statement that soft and collinear gluons do not flip quark helicities. There may be an issue here, however, 
as to what 
one means {\it experimentally} by a soft gluon. If, for example, a gluon emitted from an out-going $b$ quark with a relative 
$p_T \sim 4$ GeV is sufficient to violate the selection rules, and yet cannot be resolved experimentally, then what is the
resulting change in the background estimates? The results will clearly 
depend to some extent on the jet algorithms used and the experimental 
resolution. We intend to address this issue in a future publication \cite{coxpilkington}. 

The mass acceptance and resolution of the forward proton detectors is also a crucial issue, which depends on many factors 
including the LHC beam optics, the 
distance of the detectors from the interaction point, the closeness of the active region of the detectors to the beams, 
and the accurate knowledge of the relative positions of the detectors to the beams. De Roeck et. al. consider the case in 
which detectors are placed at 420m from the interaction point. This position is simply the distance 
from the interaction point, with 
standard LHC beam optics, that protons which loose transverse momentum $m_H/2 \sim 60$ GeV emerge at least $10 \sigma$ from 
the beam. 
This large distance raises a serious issue. Without modification of the level 1 
trigger systems of ATLAS and CMS, the light travel time from 420m detectors is very close to, and possibly larger than, the 
time required for a level 1 trigger decision. 
This means that a trigger strategy based on the central detectors alone may be required, at least until
the proton tagger information becomes available at level 2. For dijets of such low transverse momentum ($\sim 60$ GeV), 
this is certainly a 
challenge. Both De Roeck et. al. and Boonekamp et. al. \cite{Boonekamp:2004nu} consider some basic ideas based on the 
central system topology, but it is fair to say that much work still needs to be done in this area. 

The resolution of the detectors is a crucial number. The signal to background 
$S/B \propto \Lambda(H\rightarrow gg)/\Delta M \propto G_F M^3_H / \Delta M$, where $\Delta M$ is the mass window within which 
the search is performed. This is easily seen: a search using this technique is simply a counting experiment within a mass window,
 and since the tagger resolution will always be greater than the Higgs width, the worse the resolution the 
more continuum background will enter. The achievable resolution has recently been the cause of some controversy in the 
literature \cite{Boonekamp:2004nu}, and the figures used in De Roeck et. al. may be difficult to achieve in practice \footnote{current estimates 
yield signal to background ratios of order unity for the standard model 120 GeV Higgs, rather than 3, as quoted in \cite{DeRoeck:2002hk}}
 although again more 
work needs to be done in this area.         
         
It has recently been 
suggested that the Standard Model Higgs decay to $WW^*$ may also be an interesting central exclusive channel
 \cite{KhozeDeRoeck}. Certainly in the semi-leptonic channel, such events will be kept by the CMS and ATLAS 
level 1 triggers without input from the 420m proton taggers. The hit in branching ratio at low Higgs masses 
relative to the b-quark channel is compensated for to some extent by the increased detection efficiency and 
lower backgrounds. One of the key advantages of central exclusive production in this case is that the mass 
resolution (from the proton taggers) is of course unaffected by the presence of the final state neutrino. 
And of course, observation of even a few clean events with two proton tags will provide a direct measurement of the Higgs quantum numbers.     
   
Finally, we briefly review two other scenarios in which forward proton tagging may be of significant interest at the LHC. 
Firstly, the 'intense coupling' regime of the MSSM. This is a region of MSSM parameter space in which the couplings of 
the Higgs to the electroweak gauge bosons are strongly suppressed, making discovery challenging at the LHC by conventional 
means. The rates for central exclusive production of the two scalar MSSM Higgs bosons can be enhanced by an order of magnitude
in these models, however, leading to predicted signal to background ratios in excess of 20 for masses around 
130 GeV\cite{Kaidalov:2003ys}. This region of 
parameter space can also be problematic in conventional search channels because the masses of the three neutral Higgs Bosons
are close to each other. Central production can help disentangle the Higgs bosons because, due to the spin/parity selection rules,
production of the pseudo-scalar (A) Higgs is heavily suppressed. This 
means that the 'double tagged' sample will be almost pure scalar. 

As a second example, Higgs sectors with explicit CP-violation are also an area in which central production may prove 
extremely attractive. It was also noted in \cite{Khoze:2004rc} that explicit CP violation in the Higgs sector will show itself directly as a (potentially sizeable) 
 asymmetry in the azimuthal distribution 
of the tagged protons. This measurement is probably unique at the LHC, although little detailed  phenomenological 
work has been done so far.
One such model, known as the CPX scenario \cite{Carena:2000ks}, has been shown to lead to very light (less 
than 60 GeV) Higgs bosons which would have evaded detection at LEP, and may well evade detection at the Tevatron or 
LHC \cite{Carena:2002bb}. The central production cross sections for the lightest CPX Higgs are relatively large at low 
masses \cite{Cox:2003xp}, although the acceptance of the 420m pots and trigger issues would probably rule out detection of 
light CPX Higgs 
bosons. Models other than CPX with sizeable CP violation have not been considered, however, and may have larger 
central exclusive cross sections at large enough Higgs Boson masses. Further work is certainly needed in this area.

\section{Summary}

The installation of proton tagging detectors in the 420m region around ATLAS and / or CMS would certainly add unique capabilities 
to the existing LHC experimental program. If the current calculations of central exclusive production rates survive the experimental 
tests at the Tevatron, then there is a very real chance that new particle production could be observed in this channel. For the 
Standard Model Higgs, this would amount to a direct determination of its quantum numbers, with an integrated luminosity of order 
30 fb$^{-1}$. For certain MSSM scenarios, the tagged proton channel may be the discovery channel. At higher luminosities, proton tagging 
may provide direct evidence of CP violation within the Higgs sector. There is also a potentially rich, more exotic physics menu which we 
have not discussed, including gluino and squark production, gluinoballs, and indeed any object which has $0^{++}$ or $2^{++}$ quantum 
numbers and couples strongly to gluons \cite{Khoze:2001xm}. Given the relatively low cost of such a project, and the potentially unique 
access to new physics, we believe the installation of 420m proton detectors at LHC should be given careful consideration.       

\section*{Acknowledgments}
We would like to thank Valery Khoze, Misha Ryskin, Jeff Forshaw and Albert DeRoeck for many useful discussions. This work was
funded by PPARC.

\end{document}